\documentstyle[12pt,epsf]{article}

\newcommand{\abstracts}[1]{{
\centering{\begin{minipage}{12.2truecm}
\normalsize\baselineskip=15pt
\centerline{\footnotesize ABSTRACT}\vspace*{0.3cm}
\parindent=20pt #1
\end{minipage}}\par}}

\newcommand{\eq}[1]{(\ref{#1})}
\newcommand{\diff}{\partial}
\newcommand{\beqn}{\begin{eqnarray}}
\newcommand{\eeqn}{\end{eqnarray}}
\newcommand{\cD}{{\cal D}}
\def\cC{{\cal C}}
\newcommand{\dD}{{\cal D}}

\def\dd{{\rm d}}
\def\NP{Nucl.~Phys.}

\begin{document}

\pagestyle{myheadings}
\thispagestyle{empty}
\markright{JETP Lett., Vol. 68, No. 2, 25 July 1998, pp.117-123.}
\begin{flushright}
{\large ITEP-TH-52/98}
\end{flushright}
\vspace{0.3cm}
\begin{center}
{\Large \bf String representation of $\mathbf{SU(3)}$ gluodynamics in
the abelian projection}

\vspace{0.7cm}

{\large M.~N.~Chernodub$^{\,\mbox{a}}$ and D.~A.~Komarov$^{\,\mbox{a,b}}$ 
}\\

\vspace{.5cm}

$^{\,\mbox{a}}$
{\sl Institute of Theoretical and Experimental Physics,}\\
{\sl 117259, Moscow, B.Cheremushkinskaya, 25}

\vspace{0.3cm}

$^{\,\mbox{b}}$
{\sl Moscow Institute of Physics and Technology,} \\
{\sl 141700, Moscow region., ~Dolgoprudny}

\end{center}

\vspace{1.0cm}

\abstracts{
A dual Ginzburg-Landau model corresponding to $SU(3)$
gluodynamics in abelian projection is studied. A string theory describing
QCD string dynamics is obtained in this model. The interaction of static
quarks in mesons and baryons is investigated in an approximation to
leading order.}

\vspace{.5 cm}
\vspace{1cm}

One approach to the problem of color confinement in
quantum $SU(N)$ gluodynamics is the method of abelian projections,
proposed by 't Hooft \cite{tHo81}.
This method is based on partial gauge fixing, which does not fix the
abelian gauge
subgroup ${\bigl[U(1)\bigr]}^{N-1}$.
The diagonal elements of the gluon field transform under such abelian
transformations as gauge fields, while the off-diagonal elements transform
as  matter vector fields.
Since $SU(N)$ group is compact, its abelian subgroup is also compact,
and abelian monopoles exist in the system. 
If the monopoles are condensed, then confinement can be explained at
the classical level \cite{Man76,tHo76}: a string forms between
color charges (quarks), which is a (dual) analog of the Abrikosov
string~\cite{Abr} in a superconductor, the role of the Cooper pairs
being played by the monopoles.

The confinement mechanism described above, which is often called the
"dual superconductor mechanism," has been confirmed by numerous 
computer calculations on a lattice (see, for example, the reviews
in \cite{RecentReviews}). In particular, it has been shown
 that 
the contribution of abelian monopoles to the tension of the string
is almost completely identical to the total $SU(2)$ string
tension~\cite{AbelianDominance},
the monopole currents satisfy the London equation for a
superconductor~\cite{LondonEquation},
and the condensate of abelian monopoles is different from zero in the 
confinement phase and strictly equals zero in the deconfinement
phase~\cite{Condensate}.
Although these results were all obtained for 
$SU(2)$ gluodynamics,
one can believe
that the dual superconductor model should 
also work
in the more realistic case of
$SU(3)$ gluodynamics \cite{RecentReviews}.
A substantial difference of the $SU(3)$ theory from
$SU(2)$ theory
is the presence of two independent string configurations --- 
dual analogs of the Abrikosov string in a superconductor.
In the present letter we study
the string degrees of freedom and 
investigate interaction of quarks in 
$SU(3)$ gluodynamics on the
basis of the dual superconductor model (the dual Ginzburg-Landau model).

In Euclidean space, the dual Ginzburg-Landau model corresponding to 
$SU(3)$ gluodynamics is given by the lagrangian~\cite{suzuki}
\beqn
{\cal L}_{\rm DGL}={1 \over 4}(\partial_\mu
\vec B_\nu - \partial_\nu  \vec B_\mu  )^2 +\sum_{i =1}^3
\Bigl[|(i\partial_\mu -g \vec \epsilon_i \cdot \vec B_\mu)\chi_i |^2
+\lambda (|\chi_i |^2-v^2)^2\Bigr]\,.\label{ldgl}
\eeqn
This lagrangian contains two abelian gauge fields,
$\vec B_\mu = (B^3_\mu,B^8_\mu)$, 
which are dual to the gluon fields
$A^3$ and $A^8$, belonging to the Cartan subgroup of the 
$SU(3)$ gauge group. The model \eq{ldgl} also contains three Higgs fields 
$\chi_k = \rho_k \, e^{i \theta_k}$, $k=1,2,3$, 
and the phases $\theta_k$
are related by condition
\beqn
\theta_1 + \theta_2 + \theta_3 = 0\,.
\label{relation}
\eeqn
The Higgs fields  $\chi_i$ 
correspond to the monopole fields, the monopoles being condensed, since
 $\lambda>0$ and $v^2 >0$. The abelian charges of the Higgs
 fields with respect to the gauge fields
 $B^3_\mu$ and $B^8_\mu$
are determined by the root vectors of the group $SU(3)$:
$\vec \epsilon_1=(1, 0)\,,\vec \epsilon_2=(-1 \slash 2,
-{\sqrt{3} \slash 2})\,,\vec \epsilon_3=(-{1 \slash 2}, {\sqrt{3}
\slash 2})$; we also used the notation
 ${({\vec a})}^2 = ({\vec a},{\vec a})$, 
where $({\vec a},{\vec b}) = a^3 b^3 + a^8 b^8$.
The lagrangian \eq{ldgl} is invariant under $[U(1)]^2$
gauge transformations:

$$
\begin{array}{ccll}
B^a_\mu & \to & B^a_\mu + \partial_\mu \alpha^a\,, &
\quad a=3,8\,;\\
\theta_i & \to & \theta_i + g (
{\vec \epsilon}_i,{\vec \alpha})\,\, \mathrm{
mod \, 2 \pi}\,, & \quad i =1,2,3\,;\\
\end{array}
$$
where ${\vec \alpha} = (\alpha^3,\alpha^8)$ are the parameters of
the gauge transformation.

The model \eq{ldgl} contains vortex configurations,
which are analogous to the Abrikosov--Nielsen--Olesen (ANO) strings
\cite{Abr,NO} in the Abelian Higgs Model (AHM).
In the AHM, the electrically charged Higgs fields $\Phi$ are condensed
and the ANO strings carry a quantized magnetic flux.
In a circuit around such a string the phase $\theta = \arg \Phi$ 
of the Higgs field
acquires an increment
 $\theta \to \theta + 2 \pi n$,
  where $n$ is an integer (the number of elementary fluxes
  inside the string).
 Thus, at the center of the string the phase of the Higgs field is singular,
  and therefore the Higgs field equals zero:
   ${\rm Im} \Phi
= {\rm Re} \Phi = 0$. 
These last two equations define a two-dimensional manifold
in a four-dimensional space, which is the world surface
of the center of the string.

Three string degrees of freedom $\Sigma^{(i)}$ which correspond
to three Higgs fields $\chi_i$,
$i=1,2,3$
can be introduced in the model \eq{ldgl}.
Since in this model the condensed Higgs fields possess magnetic charge,
the strings $\Sigma^{(i)}$ carry an electric flux and, as we shall see
below, they possess nonzero tension, which results in color confinement.
By analogy with the Abelian Higgs Model \cite{AhChPoZu96} the strings
$\Sigma^{(i)}$ are determined by the equations
\beqn
\diff_{[\mu,} \diff_{\nu]} \theta_i (x, \tilde x) & = &
2 \pi {\tilde \Sigma}^{(i)}_{\mu\nu}(x, \tilde x^{(i)})\,, \quad
{\tilde \Sigma}^{(i)}_{\mu\nu} = \frac{1}{2}
\epsilon_{\mu\nu \alpha \beta} \Sigma^{(i)}_{\alpha \beta}\,,
\label{theta}\\
\Sigma^{(i)}_{\alpha \beta}(x, \tilde x^{(i)}) & = &
\int_{\Sigma^{(i)}} \dd^2 \sigma \epsilon^{ab}
\frac{\diff \tilde x^{(i)}_{\alpha}}{\diff \sigma_a}
\frac{\diff \tilde x^{(i)}_{\beta}}{\diff \sigma_b}
\delta^{(4)} [x - \tilde x^{(i)}(\sigma)]\,,
\nonumber
\eeqn
where the vector $\tilde x = \tilde x^{(i)}(\sigma)$ describes
the singularities of the phases 
$\theta_i$ 
parameterized by $\sigma_1$ and
$\sigma_2$, and the tensor $\Sigma^{(i)}_{\mu\nu}$
determines the position of the singularities. We note that
$\diff_{[\mu} \diff_{\nu]} \theta_i
\neq 0$, since the phases $\theta_i$ 
are singular functions.

The string world surfaces 
 $\Sigma^{(i)}$
 are not independent, since the phases of the Higgs fields are connected by
 relation \eq{relation}:
\beqn
\Sigma^{(1)}_{\mu\nu} + \Sigma^{(2)}_{\mu\nu}
+ \Sigma^{(3)}_{\mu\nu} = 0\,.
\label{cond}
\eeqn

According to numerical estimates
\cite{suganuma2}, the parameter
$\lambda$ in the lagrangian of the dual Ginzburg--Landau model
\eq{ldgl} is quite large, $\lambda \sim 65 \gg 1$.
For this reason 
we consider below the string degrees of freedom in the London
limit $\lambda\rightarrow\infty$,
which corresponds to the leading approximation in
$\lambda^{-1}$ expansion.

In the London limit, the radial degrees of freedom of the Higgs
fields are frozen on their vacuum values, 
 $\chi_i = v$.
Therefore the dynamical variables are the phases
$\theta_i$ of the Higgs fields
and the dual gauge fields 
$B^{3,8}$.
Then the path integral of the model is given by
\beqn
{\cal Z} = \int\limits_{- \infty}^{+ \infty} \dD B
\int\limits_{- \pi}^{+ \pi} \, \dD \theta_i \,
\exp\Bigl\{ - \int \dd^4 x \, {\cal L}_{\rm DGL} (B,\theta)
\Bigr\}\,,
\eeqn
where
\beqn
{\cal L}_{\rm DGL} (B,\theta) = {1 \over 4}(\partial_\mu
\vec B_\nu - \partial_\nu  \vec B_\mu  )^2 + v^2
\sum_{i=1}^3(\partial_\mu \theta_i +g \vec
\epsilon_i \cdot \vec B_\mu)^2 \,.
\label{ldgllond}
\eeqn

Integrating over the fields
$\vec B_{\mu}$ and $\theta^{(i)}$, in the manner of
\cite{AhChPoZu96,amh}, we obtain
\beqn
{\cal Z} & = & \int \cD {\mathbf \Sigma}_{\mu\nu} \,
\delta\Bigl(\sum\limits^3_{i=1}\Sigma^{(i)}_{\mu\nu}\Bigr)
\exp\Bigl\{ - S_{str} ({\mathbf \Sigma}) \Bigr\}\,,
\label{streff0}\\
S_{str} ({\mathbf \Sigma}) & = & 2
\pi^2 v^2 \int{\dd^4x \, \dd^4y\,
\sum\limits_{i=1}^3 \Sigma^{(i)}_{\mu\nu}(x)
 \cD_{m_B}(x-y) \Sigma^{(i)}_{\mu\nu}(y)}\,,
\label{streff}
\eeqn
where $m_B^2 = 3 g^2 v^2$ is the mass of the gauge fields
 $B^3$ and $B^8$,
$\cD_{m_B}$ is the propagator of the massive field: $(\Delta + m^2_B)
\cD_{m_B} (x) = \delta^{(4)} (x)$, 
and we have introduced the notation for the string variables
${\mathbf
\Sigma} = (\Sigma^{(1)},\Sigma^{(2)},\Sigma^{(3)})$.
The integration measure $\cD {\mathbf \Sigma}$ contains
the Jacobian \cite{AhChPoZu96} of the transformation from the field variables
 ($B^a_\mu,\theta_i$) to the string
variables ~(${\mathbf \Sigma}$).

It is useful to rewrite the string action 
\eq{streff}
in terms of the independent string variables  
$\Sigma^{(1)}$ and $\Sigma^{(2)}$, 
using relation \eq{cond}:
\beqn
S_{str} ({\mathbf \Sigma}) & = &
4 \pi^2 v^2 \int{\dd^4 x \dd^4y \Big\{
\Sigma^{(1)}_{\mu\nu}(x) \cD_{m_B} (x-y)
\Sigma^{(1)}_{\mu\nu}(y)} \nonumber\\
& & + \Sigma^{(1)}_{\mu\nu}(x)
\cD_{m_B}(x-y) \Sigma^{(2)}_{\mu\nu}(y)
+ \Sigma^{(2)}_{\mu\nu}(x) \cD_{m_B} (x-y)
\Sigma^{(2)}_{\mu\nu}(y) \Big\}\,.
\eeqn
In this formulation one can see that
the model contains two types of strings, which repel 
one another when 
the electric fluxes in them are parallel
 and attract one another when the fluxes are 
antiparallel.

An interesting problem is to find the interaction potential
of quarks at rest (infinitely heavy quarks).
Since the Ginzburg-Landau model under study 
is dual to $SU(3)$ gluodynamics,
the interaction potential 
 $V_c (R)$ of quarks 
  $q_c$ and $q_{{\bar c}}$
   is determined by the average of the 't Hooft
 loop:
\beqn
V_c(R) & = & - \lim\limits_{T \to + \infty}
\frac{1}{T} \, {\mathrm {ln}} <H_c(\cC_{R \times T})>\,,
\label{Vc}\\
<H_c(\cC)> & = &{1 \over {\cal Z}} \int\limits_{- \infty}^{+ \infty}
\dD B \int\limits_{- \pi}^{+ \pi} \, \dD \theta_i \, \exp\Bigl\{
- \int \dd^4 x \, \Bigl[ {1 \over 4}(\partial_\mu
\vec B_\nu - \partial_\nu  \vec B_\mu - {\vec Q}^{(c)}
\Sigma^\cC_{\mu\nu})^2 \nonumber\\
 & & + v^2 \sum_{i=1}^3(\partial_\mu \theta_i +g \vec
\epsilon _\alpha \cdot \vec B_\mu)^2 \Bigr] \Bigr\}
\,,\label{thoft}
\eeqn
where the contour $\cC_{R \times T}$ is a rectangular loop of size
$R \times T$ , representing the trajectories of 
the quark and antiquark.
The surface $\Sigma^\cC_{\mu\nu}$ is the string whose boundary
 is the trajectory $\cC$:
\beqn
\partial_\mu \Sigma^{\cC}_{\mu\nu} (x) = j^\cC_\nu (x)\,, \quad
j^\cC_\mu (x) = \oint\limits_\cC \dd \tau \frac{\partial {\tilde
x}_\mu (\tau)}{\partial \tau} \, \delta^{(4)} (x - {\tilde x}
(\tau))\,,
\eeqn
where the vector $\tilde x_\mu$ parameterizes the trajectory $\cC$. 
The quark $q_c$
(antiquark $q_{{\bar c}}$) carries color charges
${\vec Q}^{(c)}$, (${\vec Q}^{({\bar c})} = - {\vec Q}^{(c)}$,
respectively), which take the values:
\beqn
{\vec Q}^{(c)} =
(Q^{(c)}_{3},Q^{(c)}_{8})=\{({e \over 2},{e \over
2  \sqrt{3}}),\,(-{e \over 2},{e \over2 \sqrt{3}}),\, (0,-{e \over
\sqrt{3}})\}\,,
\eeqn
for red ($c=R$), blue ($c=B$), and green ($c=G$) quarks,
respectively. Here  $e=4 \pi \slash g$ is an elementary abelian electric
charge.

Integrating in expression \eq{thoft} over the fields $\vec B_\mu$ and
$\theta_i$, we obtain:
\beqn
 <H_c(\cC)> = {1 \over {\cal Z}}
\int \cD{\mathbf \Sigma}^{\cC}_{\mu\nu} \,
\delta\Bigl(\sum\limits^3_{i=1}\Sigma^{(i)}_{\mu\nu}\Bigr) \,
\exp\Bigl\{- \pi^2 v^2 \int \dd^4x \, \dd^4y &
\nonumber\\
 \sum\limits_{i=1}^3\Bigl(
\Sigma^{\cC,(i)}_{\mu\nu}\,(x;s^{(c)}_i) \cD_{m_B}(x-y)
\Sigma^{\cC,(i)}_{\mu\nu}\,(y;s^{(c)}_i) + \frac{2}{m^2_B}
\, j^\cC_\mu (x;s^{(c)}_i) \cD_{m_B}(x-y) j^\cC_\mu (y;s^{(c)}_i) \Bigr)\,,
\label{general}
\eeqn
where for string surfaces of the type $i$,
 which have the contour  $\cC$ as the boundary, we have introduced a notation:
\beqn
\Sigma^{\cC,(i)}_{\mu\nu} (x; s^{(c)}_i) =
s^{(c)}_i \, \Sigma^{\cC}_{\mu\nu}(x) + \Sigma^{(i)}_{\mu\nu}(x)\,,\quad
j^\cC_\mu (x;s^{(c)}_i)=\diff_\nu\Sigma^{\cC,(i)}_{\mu\nu} (x;
s^{(c)}_i)\,. \eeqn 
These variables satisfy the relation
\beqn
\sum\limits^3_{i=1} \Sigma^{\cC,(i)}_{\mu\nu}
(x; s^{(c)}_i) = 0\,.
\label{cond2}
\eeqn
The quantities $s^{(c)}_i$ possess a simple meaning: the quark of color 
$c$ is the boundary for the 
$s^{(c)}_i$ strings of type $i$.
 If $s^{(c)}_i < 0$, then the corresponding string carries 
a negative flux. 
The quantities $s$ are presented in the Table~1,

\begin{center}
\begin{tabular}{rccc}
~\hspace{0.8cm}c&  R & B & G  \\
$s^{(c)}_1$   &  1 & -1 & 0 \\
$s^{(c)}_2$   &  -1 & 0 & 1 \\
$s^{(c)}_3$   &   0 & 1 & -1 \\
\end{tabular} \\
\vspace{0.3cm}
{Table~1}
\end{center}
according to which only colorless states can have a finite mass: if
quarks form a colored combination, then there exists a string which
carries off the flux of the color field from this configuration to
infinity.  The energy of such a string is infinite, since the string
possesses nonzero tension.

As a result of the condition
\eq{cond2}, 
 the quarks in the pairs $R-{\bar R}$, $B-{\bar B}$ and $G-{\bar G}$
   are connected to one another 
either by two oppositely
  directed strings of two different types 
  or only by any one of these strings, depending on 
the type of the string integrated out to
 remove the $\delta$--function in \eq{general}.
 This will not affect 
 the physical quantities (for example, the interaction potential of the 
 quarks), since the tension of two oppositely directed 
 strings of the different types which are "stuck together"
  will be identical to the  tension 
 of one string. Likewise, the interaction potential of the quarks
$q_c$ and $q_{{\bar c}}$ does not depend upon color $c$.
For this reason we shall find the potential between the 
$G-{\bar G}$ 
pair at rest, having integrated over the string $i=3$ in
\eq{general}. We get:
\beqn
<H_{c=G}(\cC)> & = & {1\over{\cal Z}}
\int\cD\Sigma^{(1)}_{\mu\nu}\cD\Sigma^{\cC,(2)}_{\mu\nu}
\exp\Bigl\{- 2 \pi^2 v^2 \int \dd^4x \, \dd^4y \,
\nonumber\\
 & & \Bigl(\Sigma^{(1)}_{\mu\nu}\,(x)
\cD_{m_B}(x-y) \Sigma^{(1)}_{\mu\nu}\,(y)
+ \Sigma^{(1)}_{\mu\nu}\,(x)
\cD_{m_B}(x-y) \Sigma^{\cC,(2)}_{\mu\nu}\,(y)
\label{thoftint} \\
& & + \Sigma^{\cC,(2)}_{\mu\nu}\,(x)
\cD_{m_B}(x-y) \Sigma^{\cC,(2)}_{\mu\nu}\,(y)
+ \frac{2}{m^2_B} \,
j^\cC_\mu (x) \cD_{m_B}(x-y) j^\cC_\mu (y) \Bigr) \Bigr\}\,,
\nonumber
\eeqn
where the string world surface $\Sigma^{(1)}_{\mu\nu} (x) \equiv
\Sigma^{\cC,(1)}_{\mu\nu}\,(x;0)$ is closed, while
the string $\Sigma^{\cC,(1)}_{\mu\nu}\,(x) \equiv
\Sigma^{\cC,(1)}_{\mu\nu}\,(x;1)$ has as its boundary
 the quark-antiquark trajectory.

It is well known, that 
in the limit of large mass $M_H$ of the scalar particle
the
tension $\sigma$ of an ANO string
becomes large 
($\sigma$ increases as ${\mathrm {ln}} M_H \sim {\mathrm {ln}} \lambda$). 
Therefore it can be expected that in the London limit
the potential $V_c (R)$ is determined by the minimum of the action in
\eq{thoftint}. This minimum is reached 
on a configuration
in which the 
 string of type $i=1$ is absent and the world surface of the string of type
$i=2$ forms a minimal surface stretched on the contour $\cC$:
\beqn
\Sigma^{(1)} (x) = 0\,, \quad
{\Sigma}^{\cC,(2)} (x) =
\delta(x_2)\,\delta(x_3)\,\theta(x_1)\,\theta(R-x_1)\,,
\label{minconf}
\eeqn
the quarks being at rest at the points (0,0,0) and (R,0,0).
Substituting expressions \eq{minconf} into \eq{thoftint} and using \eq{Vc}
we obtain in the momentum representation 
(the projection of the momentum on an axis
connecting the quarks is denoted by $p_1$):
\beqn
V_{cl}(R) = - \frac{2 e^2}{3}
\int {\dd^3p \over (2 \pi)^3}\sin^2({p_1R\over 2})\Big[{1 \over
p^2+m^2_B}+{m^2_B \over p^2+m^2_B}{1 \over p_1^2}\Big] \label{potenc}\,.
\eeqn
This expression 
is distinguished by a numerical factor from the expression
 obtained in
\cite{GuPoZa98}
for the interaction potential of static quarks
in the $U(1)$ model, corresponding to the abelian projected  
$SU(2)$ gluodynamics.
The first term in \eq{potenc} corresponds to 
an exchange 
of a massive vector boson and leads only to
Yukawa potential;
the second term is of string origin.
Proceeding similarly to 
\cite{GuPoZa98}, we obtain (dropping an inconsequential additive constant)
\beqn
V_{cl}(R) & = & - \frac{e^2}{12 \pi }\Bigl\{{e^{-m_B r} \over r} +
r m_B^2 {\mathrm{ln}} ( \frac{m_\chi}{m_B}) + 4 m_B e^{- m_B
r}+4 r m_B^2 {\mathrm{Ei}}(- m_B r)\Bigr\} \,,\label{pts}
\eeqn
where $\mathrm{Ei}(x)$ is an exponential integral function
$\mathrm{Ei}(x) =  - \int^{+\infty}_{-x} \frac{e^{-t}}{t}
\dd t$.
In equation \eq{pts} we cut off the diverging integral at energies
$p^2 \sim m_\chi^2=2 \lambda v^2$, 
making the assumption that $\lambda$ is finite (but large),
which corresponds to taking 
into account the finite size of the string core~\cite{AhChPoZu96,amh}. 
At small distances ($r \ll m^{-1}_B$) this potential is 
of Coulomb type,
 while at large distances($r \gg m^{-1}_B$) it is linearly increasing
 and therefore leads to quark confinement.
We note that the coefficient of the term which grows linearly at 
large distances (the string tension) is identical to the result   
obtained in \cite{suganuma} by a different method.

The representation introduced above makes it possible
 to analyze the case of a bound state
of three quarks. In accordance with the Table 1, they are connected
by strings of all three types, as shown in Fig. 1.

The arrangement of the strings was chosen so that they would satisfy
the condition \eq{cond2}. 
If the quarks are located at the vertices of an equilateral triangle,
then the configuration 
where the point A lies at the center of the triangle
gives a minimum of the energy.
This is easy to see by integrating the
$\delta$-function in \eq{general} and obtaining
for the hadron 
 the string action in
the form \eq{streff}.
In this representation the quarks 
in a hadron
are connected by only two of the three
strings shown in Fig. 1. Thus, if the strings attract one another,
then they will "fuse together" on a certain segment $RA$. 
In leading approximation
the energy of the system is proportional to the sum of
the lengths of the segments $RA+BA+GA$.
Therefore the
classical configuration corresponds to the position
of the point A at the center.
This result is qualitatively in agreement with the conclusions drawn in
\cite{baryon} on the basis of numerical calculations.

In summary, the classical limit of the string representation 
introduced for 
$SU(3)$
gluodynamics in the present work 
on the basis of the dual
${[U(1)]}^2$ Ginzburg-Landau model can serve as a good approximation
for the description of quark--antiquark interaction in
$SU(3)$ gluodynamics.

~

We thank M.I.~Polikarpov for helpful remarks.
The work was supported in part by Grants 
No. 96-02-17230a and No. 96-15-96740
of the Russian Fund for Fundamental Research
and by Grants INTAS-RFBR-95-0681 and INTAS-94-0840.

\section*{Note Added}

After completing the present work we have learned about the
paper~\cite{Dima} in which the string representation of $SU(3)$
gluodynamics is also discussed.


\newpage

\section*{Figure}

\vspace{2cm}

\begin{figure}[!h]
\centerline{\epsfxsize0.48\textwidth\epsfbox{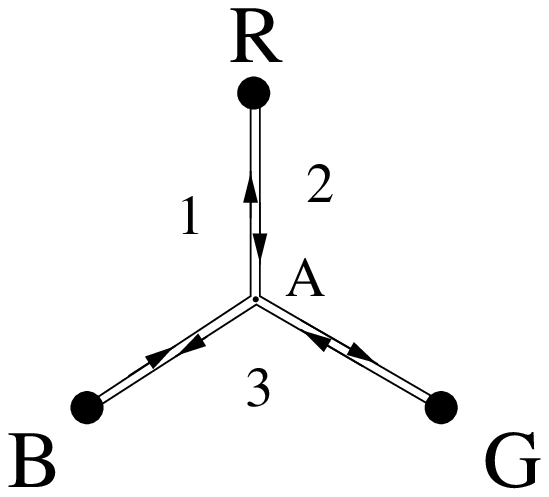}}
\vspace{1cm}
\centerline{\parbox{12cm}{Figure 1: Configuration of QCD strings
in a baryon. The letters $R$,$G$, and $B$ represent the colors of the
quarks; the numbers  enumerate
the types of strings.}}
\label{figure20}
\end{figure}

\end{document}